\documentstyle[lprocl,colordvi,epsfig,11pt]{article}
\textheight
8.5in  \parskip 7pt \openup1\jot %\parindent=0.5in
\def\st{\scriptstyle}

\def\be{\begin{equation}}
\def\ee{\end{equation}}
\def\bea{\begin{eqnarray}}
\def\eea{\end{eqnarray}}
\newskip\humongous \humongous=0pt plus 1000pt minus 1000pt
\def\caja{\mathsurround=0pt}
\def\eqalign#1{\,\vcenter{\openup1\jot \caja
        \ialign{\strut \hfil$\displaystyle{##}$&$
        \displaystyle{{}##}$\hfil\crcr#1\crcr}}\,}
\newif\ifdtup

\def\eqright #1\cr{\noalign{\hfill$\displaystyle{{}#1}$}}
\def\eqleft #1\cr{\noalign{\noindent$\displaystyle{{}#1}$\hfill}}
\def\oldreffmt#1{\rlap{[#1]} \hbox to 2\parindent{}}

\def\figfmt#1{\rlap{Figure {#1}} \hbox to 1in{}}

\def\begineq #1\endeq{$$ \refstepcounter{equation}\eqalign{#1}\eqno
	(\theequation) $$}
\def\contlimit{\,{\hbox{$\longrightarrow$}\kern-1.8em\lower1ex
\hbox{${\scriptstyle (a\rightarrow0)}$}}\,}
\def\centeron#1#2{{\setbox0=\hbox{#1}\setbox1=\hbox{#2}\ifdim
\wd1>\wd0\kern.5\wd1\kern-.5\wd0\fi
\copy0\kern-.5\wd0\kern-.5\wd1\copy1\ifdim\wd0>\wd1
\kern.5\wd0\kern-.5\wd1\fi}}
\def\centerover#1#2{\centeron{#1}{\setbox0=\hbox{#1}\setbox
1=\hbox{#2}\raise\ht0\hbox{\raise\dp1\hbox{\copy1}}}}
\def\centerunder#1#2{\centeron{#1}{\setbox0=\hbox{#1}\setbox
1=\hbox{#2}\lower\dp0\hbox{\lower\ht1\hbox{\copy1}}}}
\def\lsim{\;\centeron{\raise.35ex\hbox{$<$}}{\lower.65ex\hbox
{$\sim$}}\;}
\def\gsim{\;\centeron{\raise.35ex\hbox{$>$}}{\lower.65ex\hbox
{$\sim$}}\;}
\def\st#1{\centeron{$#1$}{$/$}}
\def\super#1{\ifmmode \hbox{\textsuper{#1}}\else\textsuper{#1}\fi}
\def\textsuper#1{\newcount\holdspacefactor\holdspacefactor=\spacefactor
$^{#1}$\spacefactor=\holdspacefactor}
\def\getcite#1,{\advance\citenumber by1
\ifnum\citenumber=1
\ref{#1}\let\next=\getcite\else\ifx#1@\let\next=\relax
\else ,\ref{#1}\let\next=\getcite\fi\fi\next}
\def\upon #1/#2 {{\textstyle{#1\over #2}}}
\relax

\def\til#1{\centeron{\hbox{$#1$}}{\lower 2ex\hbox{$\char'176$}}}
\def\tild#1{\centeron{\hbox{$\,#1$}}{\lower 2.5ex\hbox{$\char'176$}}}
\def\sumtil{\centeron{\hbox{$\displaystyle\sum$}}{\lower
-1.5ex\hbox{$\widetilde{\phantom{xx}}$}}}

\def\kbar{\underline{k}}

\def\pom{{\rm P\kern -0.53em\llap I\,}}
\def\spom{{\rm P\kern -0.36em\llap \small I\,}}
\def\sspom{{\rm P\kern -0.33em\llap \footnotesize I\,}}

\begin{document} 

\begin{titlepage} 

\rightline{\vbox{\halign{&#\hfil\cr
&ANL-HEP-CP-98-67 \cr
&\today\cr}}} 
\vspace{1.25in} 

\begin{center}
{\bf SOLVING QCD USING MULTI-REGGE THEORY}\footnote{Work 
supported by the U.S.
Department of Energy, Division of High Energy Physics, \newline Contracts
W-31-109-ENG-38 and DEFG05-86-ER-40272} 
\medskip

Alan. R. White\footnote{arw@hep.anl.gov }
\end{center}
\vskip 0.6cm

\centerline{High Energy Physics Division}
\centerline{Argonne National Laboratory}
\centerline{9700 South Cass, Il 60439, USA.}
\vspace{0.5cm}

\begin{abstract} 

This talk outlines the 
derivation of a high-energy, transverse momentum cut-off, solution
of QCD in which the Regge pole and ``single gluon'' properties of 
the pomeron are directly related to the confinement and chiral symmetry 
breaking properties of the hadron spectrum. In first approximation, the 
pomeron is a single reggeized gluon
plus a ``wee parton'' component that compensates for the color and particle
properties of the gluon. This solution corresponds to a supercritical phase
of Reggeon Field Theory. 

\end{abstract} 

\vspace{2in}
\begin{center}

Presented at the third workshop on ``Continuous Advances in QCD''
\newline University of Minnesota, Minneapolis, April 16-19, 1998.

\end{center}

\end{titlepage}

\section{Introduction}

To solve QCD at high-energy we must find both the hadronic 
states and the exchanged pomeron giving unitary scattering amplitudes.
Experimentally (in first approximation) the pomeron appears to be 
a Regge pole at small $Q^2$ and~\cite{h1} 
a single gluon at larger $Q^2$. Neither property 
is present in QCD perturbation theory.
In this talk I will outline a high-energy, transverse momentum cut-off,
``solution'' of QCD in which these non-perturbative properties of the
pomeron are directly related to the confinement and chiral symmetry breaking
properties of hadrons. 

The arguments have taken me a long time to assemble. They involve

\begin{itemize}

\item[{i)}] the techniques of multi-regge QCD calculations,

\item[{ii)}]  the dynamics of the massless quark U(1) anomaly,

\item[{iii)}] Reggeon Field Theory phase-transition analysis.

\end{itemize} 
The emphasis will be on ii) in this talk. A major outcome of the 
results~\cite{arw97} is a demonstration of how confinement and chiral
symmetry breaking, normally understood as consequences of the vacuum, can
instead be produced by a ``wee-parton'' distribution. This is a very
non-trivial property that provides, I hope, a deeper basis for the parton
model (and even~\cite{kw} the constituent quark model) in QCD ! 

The framework for my analysis is multi-regge theory. By using 
reggeon unitarity equations~\cite{gpt,arw1}, well-known Regge limit QCD
calculations~\cite{bfkl,bs,fs1} can be extended to obtain multiparticle
amplitudes involving multiple exchanges of reggeized gluons
and quarks in a variety of channels. In particular we can 
study amplitudes, such as those of the form illustrated in Fig.~1, in which
Regge pole bound states (e.g. the pion) and their scattering 
amplitudes (pomeron exchange) appear. Presently the simultaneous study of 
bound states and their scattering amplitudes is impossible in any other
formalism. 

\noindent \parbox{3.8in}{
\leavevmode
\epsfxsize=3.4in
\epsffile{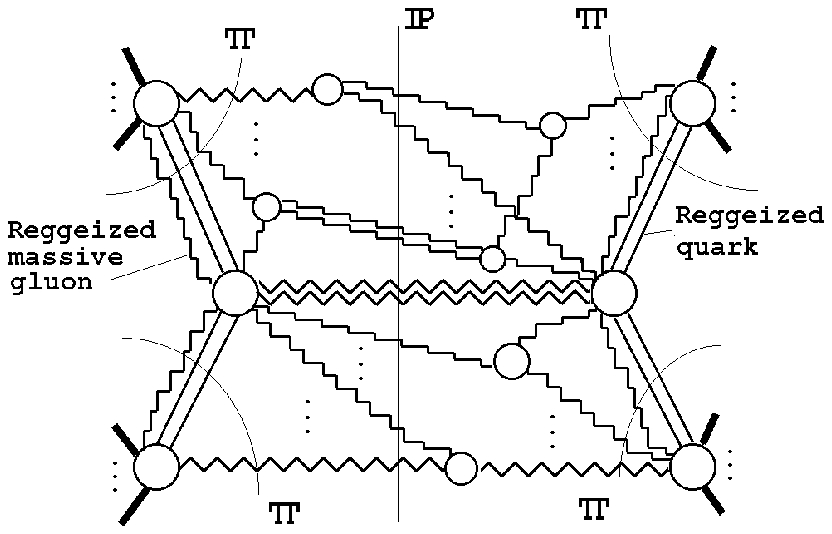}
}
\parbox{2.1in}{
\leavevmode
\epsfxsize=1.8in
\epsffile{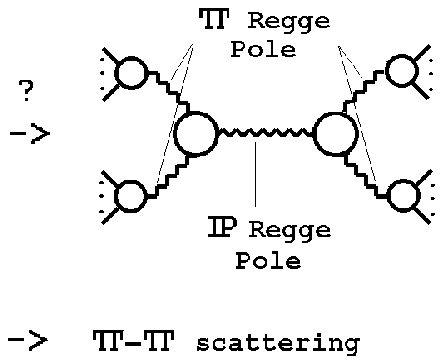}
}

\begin{center}
Fig.~1 The Anticipated Formation of Pion 
Scattering Amplitudes
\end{center}

In general, limits wrt all mass, gauge symmetry, and cut-off parameters
are crucial. The most important feature, however, is the dynamical role 
played by new ``reggeon helicity-flip'' vertices that appear in the amplitudes
we discuss. Hadron amplitudes are initially isolated via a
(``volume'') infra-red divergence that appears when SU(3) gauge symmetry is
partially broken to SU(2) and the 
limit of zero quark mass is also taken. The divergence is
produced by quark loop helicity-flip vertices involving chirality
violation (c.f. instanton interactions). 
The chirality violation survives the massless quark limit 
because of an infra-red effect closely related to 
the triangle anomaly~\cite{cg}. 
The divergence
produces (what we call) a ``wee parton condensate'' 
which is directly responsible, when the gauge symmetry is
partially broken, for confinement and chiral symmetry breaking.

The pomeron, in first approximation, is a reggeized gluon in the wee
parton condensate and so is obviously a Regge pole. Although we will not give 
any description of supercritical RFT~\cite{arw1} in this talk we do find
that all the essential features of this phase are present. We briefly discuss
the restoration of SU(3) gauge symmetry. It is closely related with the
critical behaviour of the pomeron~\cite{cri} 
and the associated disappearance of the supercritical condensate. We note 
that the large $Q^2$ of deep-inelastic scattering provides a 
finite volume constraint that can keep the theory (locally) in the
supercritical phase as the full gauge symmetry is restored. A single gluon
(in the background wee parton condensate) should then be a good
approximation for the pomeron. Finally we discuss the (very special) 
circumstances under which our solution can be realized in QCD. 

\section{Multi-Regge Theory} 

This is an abstract formalism based on the existence of asymptotic
analyticity domains for multiparticle amplitudes derived~\cite{arw1,sw}
via ``Axiomatic
Field Theory'' and ``Axiomatic S-Matrix Theory''. All the assumptions made
are expected to be valid in a completely massive spontaneously-broken gauge
theory. Since we begin with massive reggeizing gluons, this is
effectively the starting point for our analysis of QCD. We can very briefly
list the key ingredients as follows. 

\noindent {\it i)  Angular Variables } 
\newline For an N-point amplitude we can introduce 
variables corresponding to any Toller diagram, i.e. any tree diagram, drawn as 
in Fig.~2, that involves
only three-point vertices. The result 
\newline \parbox{3.2in}{ \openup\jot is that we can write 
$$
M_N(P_1,..,P_N) \equiv
M_N\left(t_1,..,t_{N-3},g_1,..,g_{N-3}\right)
$$
where $t_j=Q_j^2$ and $g_j $ is in the little group of $Q_j$, i.e. 
for $t_j > 0$, $g_j \in$ SO(3), and for $t_j < 0$, 
$g_j \in$ SO(2,1). A set of 3N - 10 independent 
variables is obtained, 
N-3 $~t_i$ variables, 
N-3 $~z_j ~(\equiv \cos\theta_j)$ variables 
and 
N-4 $~u_{jk}~ (\equiv e^{i(\mu_j - \nu_k)})$ variables.

}
\parbox{2.8in}{
\begin{center}
\leavevmode
\epsfxsize=2.3in
\epsffile{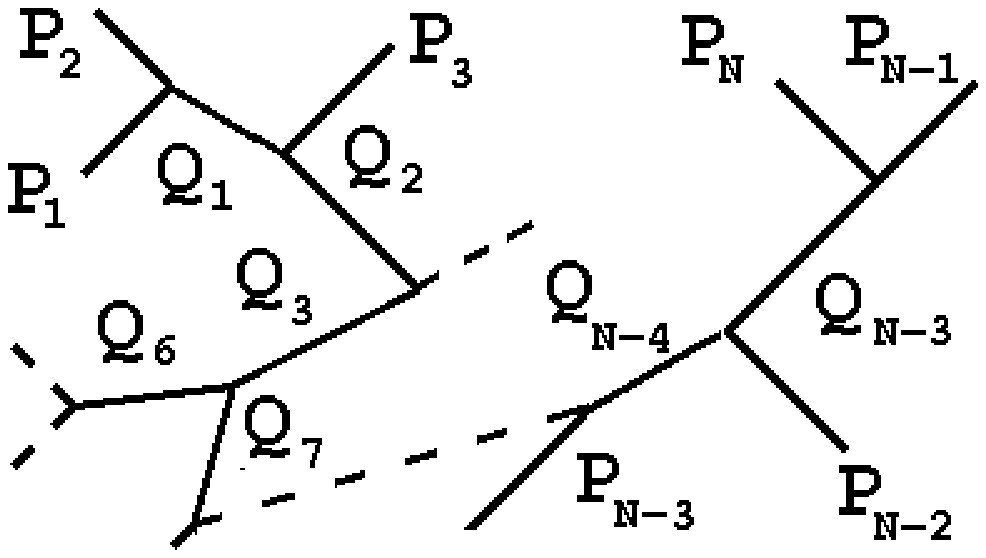}
$~$ \newline 
Fig.~2 A Tree Diagram with Three 
\newline Point Vertices.
\end{center}
}

\noindent ii) {\it Multi-Regge Limits } 
\newline These limits are defined by  
$z_j \to \infty ~,~\forall j$. We will also be interested in 
{\it Helicity-Pole Limits} in which some $u_{jk} 
\to \infty$ and some $z_j \to \infty$. In a helicity-pole limit a smaller 
number of invariants is taken large. In a ``maximal'' helicity-pole limit 
the maximal number of $u_{jk}$ are taken large.

\noindent iii) {\it Partial-wave Expansions}
\newline Using 
$ f(g)=\sum^\infty_{J=0}\,\sum_{|n|,|n'|<J}D^J_{nn'}(g)a_{J nn'}$, 
for a function $f(g)$ defined 
on SO(3), leads to 
$$
M_N(\til{t},\til{g})=\sum_{\til{J},\til{n},\til{n'}}
\prod_i~D^{J_i}_{n_in_i'}(g_i)~ a_{\til{J},\til{n},\til{n'}} (\til{t})
$$

\noindent iv) {\it Asymptotic Dispersion Relations}
\newline By dispersing in all $z_j$ variables simultaneously, and applying 
the Bargman-Weil formula, we can write $~~M_N~=~\sum_{C}M_N^C +M^0~~~~$ where
$$
M_N^C ~=~ {1\over (2\pi i)^{N-3}}
\int \frac{ ds'_1\ldots ds'_{N-3}\Delta^C(
..t_i.,..u_{jk}.,..s'_i.)}
{(s'_1-s_1)(s'_2-s_2)\ldots (s'_{N-3}-s_{N-3})} 
$$
and $\sum_{C}$ is over all sets of (N-3) Regge limit asymptotic cuts.
$M^0$ is non-leading in the multi-regge limit. The resulting separation into
spectral components, which can be described using a ``hexagraph'' 
notation~\cite{arw97,arw1}, is crucial for the development of multiparticle
complex angular momentum theory. 

\noindent v) {\it Sommerfeld-Watson Representations 
of Spectral Components} 
\newline For each spectral component a multiple transformation of the 
partial-wave expansion can be performed, e.g.
$$
\eqalign{ M^C_4=&{1\over 8}\sum_{{\scriptstyle N_1, N_2}} \int
{dn_2  dn_1 dJ_1 ~u_2^{n_2} u_1^{n_1}
d^{J_1}_{0,n_1}(z_1)
d^{n_1+N_2}_{n_1,n_2}(z_2)d^{n_2+N_3}_{n_2,0}(z_3)
\over
\sin\pi n_2\sin\pi(n_1-n_2)\sin\pi(J_1-n_1)}~a^C_{N_2N_3}(J_1,n_1,n_2,
\til{t})\cr
& ~+~~\sumtil_{\til{\scriptstyle J}\til{\scriptstyle
n}}d^{J_1}_{0,n_1}(z_1)u_1^{n_1}d^{J_2}_{n_1,n_2}
(z_2)u_2^{n_2}d^{J_3}_{n_2,0}(z_3)a_{\til{\scriptstyle J}\til{\scriptstyle
n}}(\til{t})
}
$$
These representations give the form of the asymptotic behaviour in both
multi-Regge and helicity-pole limits. In particular, 
in a ``maximal'' helicity-pole limit, 
in which the maximal number of $u_{jk} 
\to \infty$, only a single (analytically-continued) partial-wave amplitude 
appears.

\noindent vi) {\it $t$-channel Unitarity in the $J$-plane}
\newline After the hexagraph separation, multiparticle unitarity 
in every $t$-channel can be projected and continued to complex $J$ 
as an equation for partial-wave amplitudes, i.e. 
$$ 
a^+_J - a^-_J= i\int d\rho \sum_{\til{N}} 
\int {dn_1 dn_2  \over 
sin\pi(J-n_1-n_2) }\int {dn_3 dn_4 \over sin\pi(n_1 -n_3 -n_4)} ~\cdots 
~a^+_{J\til{N}
\til{n}}a^-_{J\til{N}\til{n}} 
$$
Regge poles at $n_i=\alpha_i$, together with the phase-space 
$\int d\rho $ and the ``nonsense poles'' at 
$J= n_1 +n_2 -1, n_1=n_3 + n_4 -1 , ~...~$ generate multi-reggeon 
thresholds, i.e. Regge cuts. 

\noindent vii) {\it Reggeon Unitarity }
\newline In ANY $J$-plane of any partial-wave amplitude, the ``threshold''
discontinuity due to $M$ Regge poles with trajectories 
$\til{\alpha}= (\alpha_1, \alpha_2, \cdots \alpha_M)$
is given by the reggeon unitarity equation 
$$ 
\centerunder{disc}{\raisebox{-3mm}{$\scriptstyle J=\alpha_M(t)$}}~~ 
a_{\til{N} \til{n}}(J) 
~=~ {\xi}_{M} \int d\hat\rho~
a_{\til{\alpha}}(J^+)
a_{\til{\alpha}}(J^-)
{\delta\left(J-1-\sum^M_{k=1}
(\alpha_k-1)\right)\over \sin{\pi\over 2}(\alpha_1-\tau'_1)\ldots\sin{\pi\over
2}(\alpha_M-\tau'_M) }
$$
Writing $t_i=k_i^2~~$ (with 
$\int dt_1 dt_2 \lambda^{-1/2}(t,t_1,t_2) =2\int d^2 k_1 d^2 k_2 \delta^2(k 
- k_1 - k_2) $), $\int d \hat{\rho} $ can be written in terms of
two dimensional ``$~k_{\perp}$'' integrations, 
anticipating the reggeon diagram results of 
direct $s$-channel high-energy calculations~\cite{bfkl,bs,fs1}. 
The generality of reggeon unitarity makes it particularly powerful when 
applied to the
partial-wave amplitudes appearing in (maximal) helicity-pole limits. 

\section{Reggeon Diagrams in QCD}

Leading-log Regge limit calculations of elastic and multi-regge production 
amplitudes in (spontaneously-broken) gauge theories 
show\cite{bfkl,bs,fs1} that both gluons and quarks ``reggeize'', i.e.
they lie on Regge trajectories. Non-leading log calculations are described
by ``reggeon diagrams'' involving reggeized gluons and
quarks. Reggeon unitarity implies that a complete set of reggeon
diagrams arise from higher-order contributions.

Gluon reggeon diagrams involve a reggeon propagator for each reggeon state 
and also gluon particle poles e.g. the two-reggeon state 
\newline\parbox{1.1in}{
\begin{center}
\leavevmode
\epsfxsize=0.6in
\epsffile{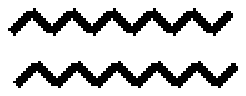}
\end{center}
}
\parbox{4in}{
$$
~\longleftrightarrow ~~~~~
\int {d^2k_1 \over (k_1^2 +M^2) } {d^2k_2 \over (k_2^2 + M^2)}~ 
{\delta^2(k_1'+k_2'-k_1-k_2)
\over J-1 +  \Delta(k_1^2) + \Delta(k_2^2)}
$$}
\newline The BFKL equation~\cite{bfkl}
corresponds to 2-reggeon unitarity, as illustrated in Fig.~3
\begin{center}
\leavevmode
\epsfxsize=4.5in 
\epsffile{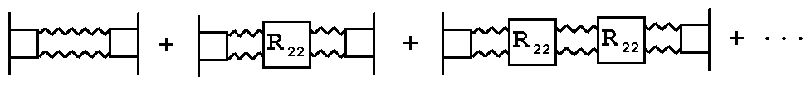}

Fig.~3 Iteration of the 2-Reggeon State.
\end{center}
with $ 
R_{22}= 
[(\kbar^2_1+M^2)({\kbar^2_2}'+M^2)+(\kbar^2_2+M^2)(
{\kbar^2_1}'+M^2)]/[ (\kbar_1-\kbar_1')^2+M^2]~ ~+ ~\cdots 
$

We assume that two leading-order properties of the limit $M \to 0$
generalize to all orders. The first is that infra-red divergences
exponentiate to zero all diagrams that do not carry
zero color in the $t$-channel. The second property is that 
infra-red finiteness implies canonical 
scaling ($\sim Q^{-2}$) for color zero reggeon 
amplitudes when all transverse momenta are simultaneously scaled to zero
(this requires $\alpha_s(Q^2) \st{\to} \infty$ when $Q^2 \to 0$).

\section{Reggeon Diagrams for Helicity-Pole Limit Amplitudes}

For our purposes, ``maximal'' helicity-pole limits of multiparticle 
amplitudes are the most interesting to study. Because the Sommerfeld-Watson
representation involves only a single partial-wave amplitude,
reggeon unitarity implies that reggeon diagrams again appear. Although we
will not discuss it here, the physical significance of such diagrams is
subtle~\cite{arw97}. In particular,  ``physical'' $k_{\perp}$ planes 
in general contain lightlike momenta ! 

As an example, 
we introduce variables for the 8-pt amplitude 
corresponding to the tree diagram of Fig.~4.
We consider the ``helicity-flip'' limit 
$ z,u_1,u^{-1}_2,u_3,u^{-1}_4 \to \infty $. The 
\newline \parbox{2.8in}{\openup\jot
behavior of invariants is
$$
\eqalign{&P_1.P_2 \sim u_1u^{-1}_2~, ~~~P_1.P_3 \sim u_1zu_3~, \cr
& P_2.P_4 \sim u^{-1}_2u^{-1}_4~, ~~~P_1.Q_3 \sim u_1z~, \cr
&Q_1.Q_3 \sim z~, ~~~P_4.Q_1 \sim zu^{-1}_4 
~ ~~ \cdots \cr
&~ P_1.Q,~P_2.Q,~ P_3.Q,~P_4.Q  ~~~\hbox{{\it finite }} }
$$
($u_1,u_2^{-1} \to \infty$ is a ``helicity-flip'' limit, 
$u_1,u_2 \to \infty$ is a ``non-flip'' limit.) }
\parbox{3.2in}{
\begin{center}
\leavevmode
\epsfxsize=2.4in
\epsffile{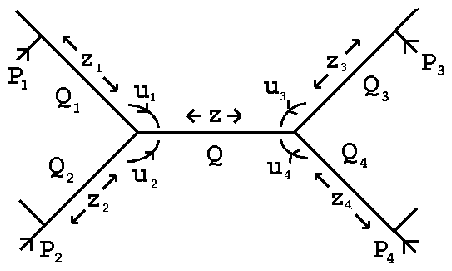}
\newline Fig.~4 Variables for the 8-pt Amplitude
\end{center}}

Reggeon unitarity determines that the helicity-flip
limit is described by reggeon 
\newline \parbox{3.1in}{ \openup\jot 
diagrams of the form shown in Fig.~5. The amplitudes 
$~{\raisebox{-2mm}{\epsfxsize=0.3in \epsffile{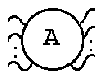}}}~$
contain all elastic scattering 
reggeon diagrams. The $T^F$ are new ``reggeon helicity-flip'' vertices
that play a crucial role in our QCD analysis. (These vertices do not appear 
in elastic scatttering reggeon diagrams).}
\parbox{2.9in}{
\begin{center}
\leavevmode
\epsfxsize=2.4in
\epsffile{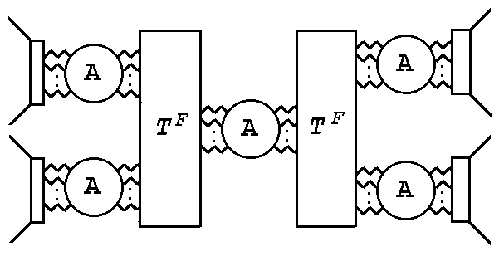}

Fig.~5 Reggeon Diagrams for 
\newline the 8-pt Amplitude
\end{center}
}

\section{Reggeon Helicity-Flip Vertices }

The $T^F$ vertices are most simply isolated kinematically by considering
a ``non-planar'' triple-regge limit which, for simplicity, we will define by 
introducing three distinct light-cone momenta. (This limit
actually gives a sum of
three $T^F$ vertices of the kind discussed above~\cite{arw97}, but in this
talk we will not elaborate on this subtlety.)  We 
use the tree diagram of Fig.~6(a) to define momenta and study the special 
kinematics 
\newline \parbox{2.5in}{
$$
\eqalign{&P_1\to (p_1,p_1,0,0),~~~p_1 \to \infty \cr
&P_2\to (p_2,0,p_2,0),~~~p_2 \to \infty \cr
&P_3\to(p_3,0,0,p_3),~~~p_3 \to \infty  \cr
&~ \cr
&Q_1\to (0,0,q_2, -q_3) \cr
&Q_2\to (0,-q_1,0,q_3) \cr
&Q_3\to (0,q_1,-q_2,0) }
$$
}
\parbox{3.5in}{
\begin{center}
\leavevmode
\epsfxsize=2.7in
\epsffile{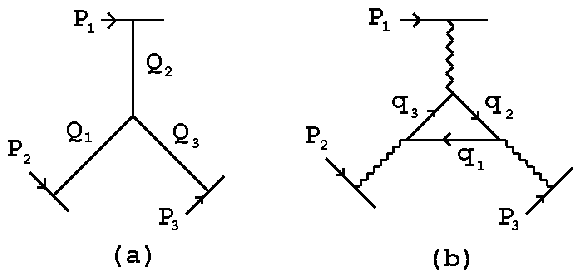}
\newline Fig.~6 (a) A Tree Diagram and (b) a quark loop coupling for three 
quark scattering.
\end{center}
}

Consider, first, three quarks scattering
via gluon exchange with a quark loop coupling as in Fig.~6(b). The 
non-planar triple-regge limit gives  
$$
\to
~ g^6~~ { p_1p_2p_3 \over t_1 t_2 t_3 } ~~\Gamma_{1^+2^+3^+}(q_1,q_2,q_3)
~~\longleftrightarrow ~~ 
g^3~~ { p_1p_2p_3 \over t_1 t_2 t_3 }~ T^F(Q_1,Q_2,Q_3) 
$$ 
where $~ \gamma_{i^+} = \gamma_0 + \gamma_i $ and 
$\Gamma_{\mu_1 \mu_2 \mu_3}$ is given by the quark triangle diagram i.e. 
$$
\Gamma_{\mu_1 \mu_2 \mu_3} = i\int {  d^4 k~ Tr \{ \gamma_{\mu_1}
(\st{q}_3 + \st{k} + m ) \gamma_{\mu_2} (\st{q_1} + \st{k} + m ) 
\gamma_{\mu_3} (\st{q}_2 + \st{k} + m) \} 
\over [ (q_1 + k)^2 - m^2 ][ (q_2 + k)^2 - m^2 ]
[ (q_3 + k)^2 - m^2 ]}
$$
where $m$ is the quark mass. We denote the $O(m^2)$ chirality-violating part of 
$~T^F ~(\equiv ~g^3 ~\Gamma_{1^+2^+3^+}~)$ by 
$T^{F,m^2}~$ and note that the 
limits $q_1, q_2, q_3 \sim Q \to 0$ and $m \to 0$ do not commute, i.e. 
$$
T^{F,m^2} {\centerunder{$\sim$}{\raisebox{-5mm} 
{$Q \to 0$} }}~Q ~i~m^2 \int {d^4k \over [ k^2 - m^2 ]^3 }
 ~~~~= ~  R ~Q 
$$
where $R$ is independent of $m$. This non-commutativity is an ``infra-red 
anomaly'' due to the triangle Landau singularity~\cite{cg}.

$T^F$ is one of 
a general set of quark loop reggeon interactions that have ultra-violet 
divergences. To maintain the reggeon Ward identities that ensure gauge
invariance~\cite{arw97}, we introduce Pauli-Villars fermions 
as a regularization. (Note that we take the regulator mass $m_{\Lambda}
\to \infty$ after $m \to 0$. This implies that the initial 
theory with $m \neq 0$ is non-unitary for $k_{\perp} \gsim m_{\Lambda}$.) 
For the regulated vertex, 
$T^{{\cal F},m^2}$, we obtain (for $m \neq 0$) 
$$
T^{{\cal F},m^2}(Q)~\sim ~ 
T^{F,m^2} ~- ~T^{F,m_{\Lambda}^2}
~~~{\centerunder{$\sim$}{\raisebox{-4mm} 
{$Q \to 0$} }} ~~~ Q^2
$$
However, since $T^{F,0} = 0$, we also have 
$$
 T^{{\cal F},0}(Q)  ~ \sim -R ~Q 
$$
implying that 
imposing gauge invariance for $m \neq 0$ gives a 
slower vanishing as $Q \to 0$ when $m=0$.

After color factors are included and all related diagrams summed, $T^{{\cal
F},0}(Q)$ survives only in very special vertices coupling reggeon states
with ``anomalous color parity''. We define color parity ($C_c$) via the
transformation $ A^i_{ab} \to - A^i_{ba}$ 
for gluon color matrices and say that a reggeon state has anomalous color 
parity if the signature $\tau$ (i.e. whether the number of 
reggeons is even or odd) is not equal to the color parity. (Note 
that the reggeized gluon 
\newline \parbox{2.7in}{ \openup\jot and the BFKL two reggeon state both have 
normal color parity.) 
We will be particularly interested in the ``anomalous odderon''
three-reggeon state with color factor
$f_{ijk}A^iA^jA^k$ that has $\tau = -1$ but $C_c = +1$ 
(c.f. the winding-number 
current
$K_{\mu}=\epsilon_{\mu \nu \gamma 
\delta}f_{ijk}A^i_{\nu}A^j_{\gamma}A^k_{\delta}$ ). $~~~T^{{\cal F},0}(Q)$ 
appears in the triple coupling of three anomalous 
odderon states as in Fig.~7. 
}
\parbox{3.3in}{
\begin{center}
\leavevmode
\epsfxsize=2.8in
\epsffile{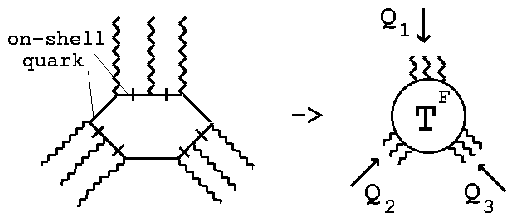}
\newline Fig.~7 An Anomalous Odderon 
\newline Triple Coupling.
\end{center}
}

\section{A Quark Mass Infra-Red Divergence} 

A vital consequence of the ``anomalous'' behavior of $~T^{{\cal F},0}~$ 
as $~Q \to 0~$ is that an 
additional infra-red divergence 
is produced (as $m \to 0$) in massless gluon reggeon diagrams. 
The divergence occurs in diagrams involving the $T^F$ 
where $~Q_1 \sim Q_2 \sim Q_3 \sim 0
~$ is part of the integration region. This requires that $T^F$ be a 
disconnected component of a vertex coupling 
\newline \parbox{2.4in}{ 
distinct reggeon channels, as in Fig.~8. 
In this diagram an anomalous odderon reggeon state ($~\equiv~$ 
$~\raisebox{-1.5mm}{$\epsfxsize=0.4in 
\epsffile{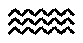}$}~$) is denoted by
$~\raisebox{-0.5mm}{$\epsfxsize=0.3in \epsffile{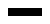}$}~$
while 
$~\raisebox{-0.5mm}{$\epsfxsize=0.3in \epsffile{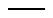}$}~$
denotes any normal reggeon state. Fig.~8 is of the general form illustrated 
in Fig.~5, except that we are allowing the vertices $V_i$ to involve more 
complicated external states than a single scattering quark. 
}
\parbox{3.5in}{ 
\begin{center}
\leavevmode
\epsfxsize=2.9in 
\epsffile{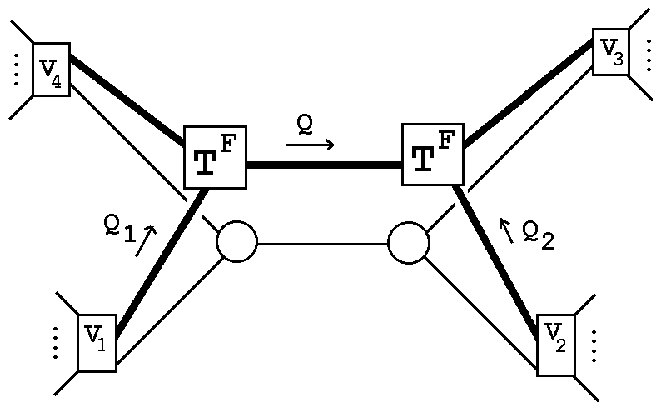}
\newline Fig.~8 A Divergent Reggeon Diagram 
\end{center}
}

The canonical 
scaling of the anomalous odderon states 
gives the infra-red behaviour 
$$
\eqalign{& \int \cdots {d^2Q_1 ~ d^2Q_2 
~d^2Q \over Q_1^2 Q_2^2 Q^2 (Q-Q_1)^2 (Q - Q_2)^2 }
~~V_1(Q_1)V_2(Q_2) V_3(Q- Q_2) V_4(Q - Q_1) \cr
&\times  ~ T^{ {\cal F}}(Q_1,Q)
T^{ {\cal F}}(Q,Q_2)
~\times~\hbox{[regular vertices and reggeon propagators]}
}
$$
for Fig.~8. Depending on the behaviour of the $V_i~$, it is clear that a
divergence may indeed occur when $Q \sim Q_1 \sim Q_2 \to 0$.
In general we can show~\cite{arw97}
that gauge invariance produces
a cancelation involving similar divergences of diagrams related to that of 
Fig.~8 by reggeon Ward identities for the reggeons within the anomalous odderon 
states. However,
the divergence of Fig.~8 is preserved and a cancelation eliminated 
if we partially break the SU(3) gauge symmetry to SU(2). 
In this case, a divergence can occur in any diagram of the form 
of Fig.~8 in which 
$~\raisebox{-0.5mm}{$\epsfxsize=0.3in \epsffile{dss15.ps}$}~$
is any SU(2) singlet 
combination of massless gluons with 
$~C_c= -\tau = +1~ $ (i.e. a generalized SU(2) anomalous odderon) and  
$~\raisebox{-0.5mm}{$\epsfxsize=0.3in \epsffile{dss16.ps}$}~$
is any normal reggeon state 
containing one or more SU(2) singlet 
massive reggeized gluons (or quarks). 

A-priori reggeon Ward identities imply $ V_i \sim Q_i$ when 
$Q_i \to 0, ~\forall ~ i $. This would actually be sufficient to eliminate any
divergence in Fig.~8. However, if we impose the ``initial condition'' that 
$V_1,V_2 ~ \st{\rightarrow}~ 0$, the divergence is present 
in a general class of diagrams, including 
\newline \parbox{2in}{
those having the general structure 
illustrated in Fig.~9. In this diagram there are 
$n + 3$ 
\newline multi-reggeon states of the form
$\raisebox{-2mm}{$\epsfxsize=0.4in \epsffile{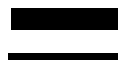}$}$ .
Imposing $V_1,V_2 ~ \st{\rightarrow}~ 0$ and assuming 
that reggeon Ward identities are satisfied by the remaining vertices, i.e. 
$$
V_i(Q_i) \sim V(Q_i) = Q_i 
$$
$i \neq 1,2$, gives that 
Fig.~9 has
}
\parbox{4in}{
\begin{center}
\leavevmode
\epsfxsize=3.6in
\epsffile{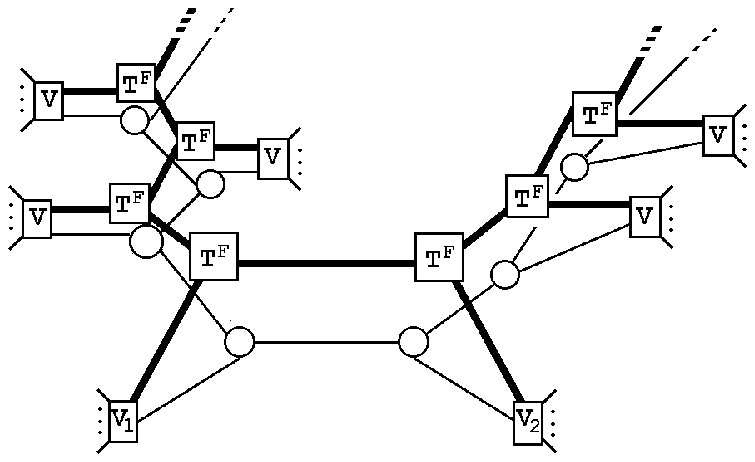}

Fig.~9 A General Divergent Diagram
\end{center}
}
\newline 
the infra-red behavior
$$ 
\int {d^2 Q \over Q^2}~\left[\int {d^2 Q \over Q^4}\right]^n 
~~\left[V(Q)~ T^ {\cal F}(Q)\right]^n 
$$
giving (as $m \to 0$) an overall logarithmic divergence. 
In general, this divergence occurs in just those multi-reggeon 
diagrams which contain only SU(2) color zero states of the form 
$~\raisebox{-2mm}{$\epsfxsize=0.4in \epsffile{cspp12.ps}$}~$
coupled by regular and $~T^{{\cal F},0}~$ vertices, as in the examples we 
have discussed.

\section{Confinement and a Parton Picture}

We define physical amplitudes by extracting the 
coefficient of the logarithmic divergence. There is ``confinement''
in that a particular set of color-zero reggeon states is selected that
contains no massless multigluon states and has the necessary completeness
property to consistently define an S-Matrix. That is, if two or more 
such states scatter 
via QCD interactions, the final states contain only 
arbitrary numbers of the
same set of states. Since $k_{\perp} =0$ for the anomalous odderon component of
each reggeon state, an ``anomalous odderon condensate'' 
is generated. 
\newline \parbox{2.2in} {
The form of physical amplitudes is illustrated in Fig.~10. 
In addition to the $k_{\perp} =0$ (``wee-parton'') component, 
each physical reggeon state has a 
finite momentum ``normal'' parton component carrying 
the kinematic properties of interactions. 
We emphasize that the ``scattering'' of the $k_{\perp} = 0$ condensate is
directly due to the infra-red quark triangle anomaly. 
}
\parbox{3.8in}{
\begin{center}
\leavevmode
\epsfxsize=3.2in
\epsffile{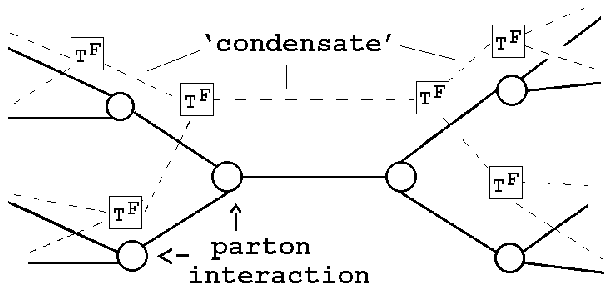}
\newline $~$
\newline Fig.~10 A Physical Amplitude
\end{center}
}

The breaking of the gauge symmetry has produced physical states in which 
the ``partons'' are separated into a universal wee-parton component and a 
normal reggeon parton component which is distinct in each distinct 
physical state. However, the condensate has the important 
property that it switches the signature compared to that of the normal parton 
component. The following are a direct consequence. 

\begin{itemize}

\item{The ``pomeron'' has a reggeized gluon normal parton component, but is a 
Regge pole with $\tau  = - C_c = + 1$ and intercept $\neq 0$. }

\item{There is a bound-state reggeon formed from two massive SU(2) doublet 
gluons, giving an exchange-degenerate partner
to the pomeron. The SU(2) singlet massive gluon lies on this trajectory. }

\item{There is chiral symmetry breaking - }
\end{itemize}
studies of $\tau = -1$ quark-antiquark exchange~\cite{ks}
can be used to demonstrate the ``reggeization 
cancelation'' shown in Fig.~11(a). 
Because of this cancelation the sum of quark-antiquark
\newline \parbox{2.2in}{
diagrams shown in Fig.~11(b) generates a Regge pole with zero intercept 
and with  
$\tau = -1, ~C_c = +1$ and $P = -1$. In the condensate we obtain a Regge 
pole with  
$\tau = +1, ~ P = -1$, giving the massless pion associated with chiral 
symmetry breaking. (We use the reggeon diagram notation that 
~\raisebox{-1mm}{$\epsfxsize=0.4in
\epsffile{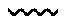}$}~ $=~$ a reggeized gluon and 
~\epsfxsize=0.3in \epsffile{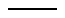}~ $=~$ a reggeized 
quark/antiquark). }
\parbox{3.8in}{
\begin{center}
\leavevmode
\epsfxsize=3in
\epsffile{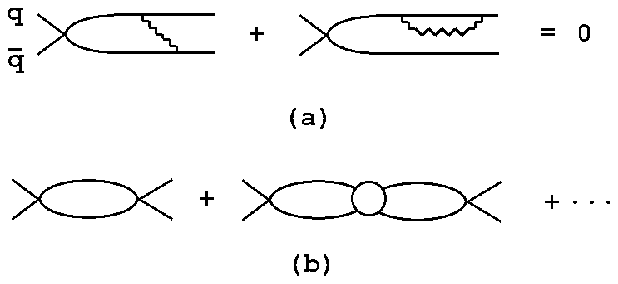}
\newline Fig.~11 (a) The Reggeization Cancelation and 
\newline (b) the Sum of Quark-AntiQuark Diagrams. 
\end{center}
}

We will not discuss Reggeon Field Theory, except to note that all the 
features of my supercritical RFT solution~\cite{arw1} are present.
(This solution was very controversial 20 years ago - although it was
supported by Gribov !) 

\section{Restoration of SU(3) Gauge Symmetry }

We make only a few brief comments on this, obviously important, subject.
To discuss it in detail requires extensive use of Reggeon Field Theory and,
in this talk, we are avoiding this. Because of complimentarity~\cite{fs},
restoring SU(3) symmetry (which involves decoupling a color triplet Higgs
scalar field) should 
be straightforward if we impose a transverse momentum cut-off 
$k_{\perp} < \Lambda_{\perp}$. Restoring the symmetry involves
removing the mass scale that produces the reggeon condensate and
distinguishes normal (finite momentum) partons from wee 
(zero momentum) partons. If the (partially) broken theory can be mapped 
completely onto supercritical RFT then the condensate 
and the odd-signature partner for the pomeron will disappear simultaneously
and the result will be the critical pomeron~\cite{cri}. The wee-parton
condensate will be replaced by a universal, small $k_{\perp}$, wee parton,
critical phenomenon that merges smoothly with the large 
$k_{\perp}$ normal (or constituent) parton component of physical states,
just as originally envisaged by Feynman~\cite{rf}. (Note that, because of
the odd SU(3) color charge parity of the pomeron, the two-gluon BFKL pomeron 
will not contribute.) 

Mapping partially-broken QCD onto supercritical RFT has further 
consequences. In particular, it implies that the 
$\Lambda_{\perp}$ scale mixes with the symmetry breaking mass scale 
and becomes a ``relevant parameter'' for the
critical behavior. It then follows that, after the symmetry breaking scale is 
removed, there will (for a general number of quark flavors) be a 
$~\Lambda_{\perp c}$ such that 
$\Lambda_{\perp} > \Lambda_{\perp c} ~$ implies the pomeron is in the 
subcritical phase, while $\Lambda_{\perp} <  \Lambda_{\perp c}~ $ will
give the 
supercritical phase. 
This implies that the supercritical phase can be
realized with the full gauge symmetry restored if $\Lambda_{\perp} $ is taken 
small enough. However, $\alpha_{\spom}(0)$ and the mass of the exchange
degenerate, composite, ``reggeized gluon'' will be functions of
$\Lambda_{\perp}$. We can also anticipate that in deep-inelastic diffraction 
large $Q^2 $ will act as an additional (local) lower $k_{\perp}$ cut-off and
produce a ``finite volume'' effect that can keep the theory supercritical as
the SU(3) symmetry is restored. 

To remove $\Lambda_{\perp}$ requires $\Lambda_{\perp c} ~= \infty $. As we 
briefly elaborate in the next Section, this requires a specific quark flavor
content. It is interesting that, for any quark content, we can take 
$\Lambda_{\perp} <<  \Lambda_{\perp c} ~  $, and go deep into the 
supercritical phase. We obtain a picture in which constituent quark 
hadrons interact via a massive composite ``gluon'' (and an exchange 
degenerate pomeron). Confinement and chiral symmetry breaking are realized 
via a simple, universal, wee parton component of physical states. This is 
remarkably close to the realization of the constituent quark model
via light-cone quantization that has been advocated by light-cone 
enthusiasts~\cite{kw}.

\section{When is this Solution Realized in QCD ??}

We have found a high-energy S-Matrix via a transverse momentum infra-red
phenomenon involving massless gluons and quarks. At first sight, it would
appear that this could not occur in QCD since non-perturbative effects
should eliminate massless gluons for $k_{\perp} < \lambda_{QCD}$! Our
solution requires that massless QCD 
remain weak-coupling at $k_{\perp} = 0$. This is the case only if there are
a sufficient number of massless quarks in the theory to 
give an infra-red fixed point for $\alpha_s$.

With the maximum number of flavors allowed by asymptotic freedom, there 
is such an infra-red fixed point and we can also break SU(3) symmetry to
SU(2) with an asymptotically-free scalar field. This can be used to show 
that $\Lambda_{\perp c} = \infty $. This, in turn, implies that critical 
pomeron scaling occurs for all $k_{\perp}$ and allows a smooth match 
with perturbative QCD. 

The above arguments suggest that if ``single gluon'' supercritical pomeron 
behavior is actually observed at HERA then new QCD physics, in the form 
of a new fermion sector, remains to be discovered above the (diffractive) 
$Q^2$ range presently covered. Everything is consistent if the electroweak
scale is a QCD scale, i.e. the ``Higgs sector'' of the Standard Model, that 
is yet to be discovered, is
composed~\cite{arw2} of higher-color (sextet) quarks. 
A special definition of QCD is necessarily involved, but we will not 
discuss this here.

\noindent { \bf References}


\begin{thebibliography}{99}

\bibitem{h1} H1 Collaboration, pa02-61 ICHEP'96 (1996), For a final version
of the analysis see {\it Z. Phys.} {\bf C76}, 613 (1997). See also 
ZEUS collaboration, hep-ex/9804013. 

\bibitem{arw97} A.~R.~White, hep-ph/9712466 (1997), to be published in Phys. 
Rev. D.

\bibitem{kw} K.~G.~Wilson, T.~S.~Walhout, A.~Harindranath, Wei-Min Zhang, 
S.~D.~Glazek and R.~J.~Perry, {\it Phys. Rev.} {\bf D 49}, 6720 (1994).

\bibitem{gpt} V.~N.~Gribov, I.~Ya.~Pomeranchuk and K.~A.~Ter-Martirosyan,
{\it Phys. Rev.} {\bf 139B}, 184 (1965).

\bibitem{arw1} A.~R.~White, Int. J. Mod. Phys. {\bf A11}, 1859 (1991);
 A.~R.~White in {\em Structural Analysis of Collision Amplitudes},
(North Holland, 1976). 

\bibitem{bfkl} E.~A.~Kuraev, L.~N.~Lipatov, V.~S.~Fadin, {\it Sov. Phys.
JETP} {\bf 45}, 199 (1977) ; Ya.~Ya.~Balitsky and L.~N.~Lipatov, {\it Sov. J.
Nucl. Phys.} {\bf 28}, 822 (1978). 
V.~S.~Fadin and L.~N.~Lipatov, {\it Nucl. Phys.} {\bf B477},
767 (1996) and further references therein.

\bibitem{bs} J.~B.~Bronzan and R.~L.~Sugar, {\it Phys. Rev.} {\bf D17}, 
585 (1978). This paper organizes into reggeon diagrams the results from 
H.~Cheng and C.~Y.~Lo, {\it Phys. Rev.} {\bf D13}, 1131 (1976), 
{\bf D15}, 2959 (1977). 

\bibitem{fs1} V.~S.~Fadin and V.~E.~Sherman, {\it Sov. Phys.} {\bf  JETP 
45}, 861 (1978).

\bibitem{cg} S.~Coleman and B.~Grossman, {\it Nucl. Phys. }
{\bf B203}, 205 (1982).

\bibitem{cri}  A.~A.~Migdal, A.~M.~Polyakov and K.~A.~Ter-Martirosyan, 
{\it Zh. Eksp. Teor.  Fiz.} {\bf 67}, 84 (1974); 
H.~D.~I.~Abarbanel and J.~B.~Bronzan, {\it Phys. Rev.} {\bf D9}, 2397 (1974).

\bibitem{sw} H.~P.~Stapp 
in {\em Structural Analysis of Collision Amplitudes},
(North Holland, 1976); 
H.~P.~Stapp and A.~R.~White, {\it Phys. Rev.} {\bf D26}, 2145 
(1982). 

\bibitem{ks} R.~Kirschner, L.~Mankiewicz and L.~ Szymanowski, {\it Z. Phys.}
{\bf C74}, 501 (1997). 
 
\bibitem{fs} E.~Fradkin and S.~H.~Shenker, {\it Phys. Rev. }
{\bf D19}, 3682 (1979); 
T.~Banks and E.~Rabinovici, {\it Nucl. Phys. } {\bf B160}, 349 (1979).
 
\bibitem{rf} R.~P.~Feynman in {\it Photon Hadron Interactions}
(Benjamin, 1972)).

\bibitem{arw2} A.~R.~White, hep-ph/9704248 (1997).

\end{thebibliography}
\end{document}